# Direct measurements of cosmic rays (TeV and beyond) in space using an ultra-thin homogeneous calorimeter

Elena Dmitrieva[1], Anastasiya Fedosimova[1,2], Igor Lebedev[1], Abzal Temiraliev[1], Ekaterina Grushevskaya[1], Sayara Ibraimova[1], Medeu Abishev[2], Tolegen Kozhamkulov[2], Andrey Mayorov[3] and Claudio Spitaleri[4]

[1] Institute of Physics and Technology, Satbayev University, Almaty, Kazakhstan
[2] Al-Farabi Kazakh National University, Almaty, Kazakhstan
[3] Moscow Engineering and Physics Institute, Moscow, Russia
[4] University of Catania, Catania, Italy

E-mail: lebedev692007@yandex.ru

**Abstract**

An approach for measuring energy of cosmic-ray particles with energies $E > 10^{12}$ eV using an ultrathin calorimeter is presented. The method is based on the analysis of the correlation dependence of the cascade size on the rate of development of the cascade process. In order to determine the primary energy, measurements are made based on the number of secondary particles in the cascade, $N_e$, at two observation levels $Z_1$ and $Z_2$, separated by an absorber layer. Based on the obtained measurements, a correlation analysis of the dependence of $\log N_e(Z_1)$ on the difference $dN = \log N_e(Z_1) - \log N_e(Z_2)$ is carried out. The correlation curves ($\log N_e$ from $dN$) in the negative part of the $dN$ axis are almost parallel to each other and practically do not depend on the depth of the cascade development. It makes it possible to determine the primary energy using an ultrathin calorimeter. The best option for applying the method is a calorimeter, which has a unit with a heavy target, leading to the rapid development of the cascade, and a homogeneous measuring and absorption block.

## 1. Introduction

Accurate knowledge of the energy spectra and elemental composition of primary cosmic rays makes it possible to understand the features of the formation of cosmic rays in astrophysical sources, and details of the processes of cosmic ray propagation in the Galaxy. Different cosmological models predict different spectra of the elements and different elemental composition of cosmic rays [1-4].

The information obtained during the study of extensive air showers at earth stations of cosmic rays is very indirect. The first interaction of a cosmic particle occurs in the upper air, forming a cascade of secondary particles. In order to reconstruct the energy and mass of the primary cosmic particle, different approaches are used, which interpret the signals in the detectors in terms of different models of hadron interactions [5-7].

Studies of the characteristics of cosmic rays based on direct measurements outside the atmosphere of Earth on spacecraft or high-altitude balloons make it possible to get more reliable information. The main advantage of direct experiments is the ability to measure the charge of an incident particle. The energies of cosmic particles are measured accurately enough for particles with energies $E<10^{11}$ eV. Modern magnetic spectrometers can determine the primary energy with an error of less than 10 percent [8-10].

Such devices have limitations at energies $E \geq 10^{12}$ eV, and the task of determining the primary energy based on direct measurements of cosmic rays becomes more complicated. Nowadays the best option for energy measurements of various nuclei in a wide energy range is the method of an ionization calorimeter [11].

The technical implementation of modern ionization calorimeters may be different, but the fundamental idea is the same. The primary particle, when entering a calorimeter and interacting with its substance, initiates the appearance of secondary particles, transferring a part of their energy to them. Secondary particles form a shower, which is absorbed in the volume of a calorimeter. In order to measure the characteristics of a cascade, a dense substance is interlaid with special detectors. Measuring the signals from these detectors the cascade curve can be constructed. The cascade curve shows the dependence of the number of particles in the cascade, $N_e$, on the depth of penetration, $d$, of the cascade in a calorimeter. If the maximum of the cascade curve is measured, then the energy of the primary particle can be determined.

The main problem with this method of measuring energy is heavy installations, since a calorimeter must have sufficient depth for plotting the cascade curve. The huge weight of the installation greatly complicates the possibility of using such a device in space experiments.

Consequently, a more promising approach for determining the energy of cosmic rays based on direct measurements is the use of a thin calorimeter. The entire cascade of secondary particles is not fixed, but only the beginning of the cascade is analyzed in a thin calorimeter. However, such measurements usually have errors of more than 50% due to the major fluctuations in the cascade development [12].

In this paper we present an approach to suppress the influence of fluctuations in the cascade development on the results of determining the primary energy.

## 2. Method of the correlation curves

To determine the primary energy $E$ based on the number of secondary particles at the observation level, usually the following dependence is used: $N_e = aE^b$, where $a$, $b$ are parameters depending on the penetration depth $d$ and the mass of the primary particle. The equation is statistically correct. However, $N_e$ strongly fluctuates in an individual event.

Figure 1 shows the shower curves initiated by iron nuclei with energies of $10^{14}$ eV, $10^{15}$ eV, $10^{16}$ eV and $10^{17}$ eV in the atmosphere of Earth. The showers were simulated using the CORSIKA QGSJET software package [13].

The figure shows that the cascade curves fluctuate substantially and practically merge (do not separate) at low values of the penetration depth $d$. For example, at a penetration depth of $d = 40$ g/cm$^2$, fluctuations are so major that some showers with an energy of $10^{14}$ eV have log $N_e$ higher than showers with an energy of $10^{15}$ eV.

What is the reason for such major fluctuations?

First of all, the number of particles in a shower depends on fluctuations of the penetration depth before the first interaction. The earlier the primary particle interacts, the more secondary particles will be at the observation level.

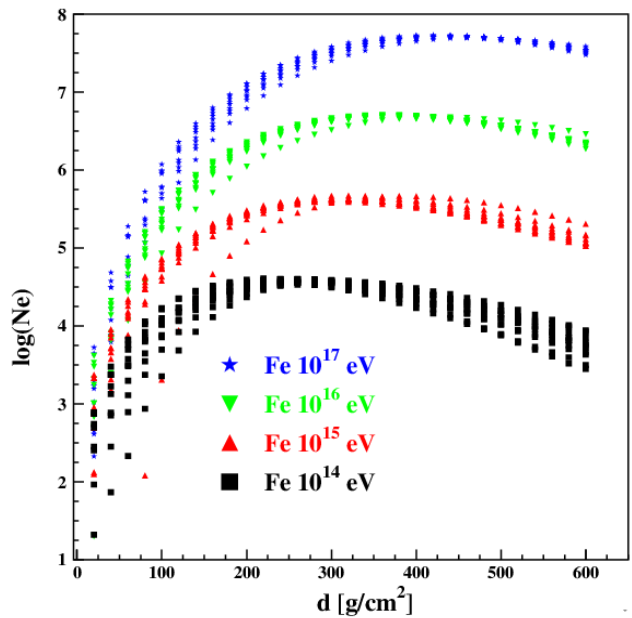

**Figure 1.** Cascade curves initiated by iron nuclei of various energies in the atmosphere of Earth.

In addition, fluctuations in the number of particles in a shower depend on the initial collision geometry and on the features of the nucleus-nucleus interaction. Different approaches and methods are used in order to study such features in the multiparticle production [14-16].

In this article we did not analyze the features of the interaction of the nucleus with the nucleus. The influence of different interaction models requires separate close examination. Therefore, we limited ourselves to a simplified representation of the presented approach. We used only the number of secondary particles as the parameter to be analyzed, since this parameter determines the rate of the cascade development. For example, showers with multiplicity in the first interaction of $M = 100$ develop faster than showers with $M = 10$.

After the first interaction, each secondary particle can interact, also producing secondary particles. When the number of secondary particles in a shower becomes large, the fluctuations of the individual interactions more or less compensate each other. Therefore, all the cascades formed by primary particles of the same mass and energy, starting from a certain moment, develop almost equally [17-18].

The situation is paradoxical one. On the one hand, major fluctuations of the cascade curve lead to large errors in determining the primary energy. On the other hand, the universal cascade development takes place, if the value of the penetration depth is not considered as a parameter. Thus, the solution to the problem is obvious. It is necessary to find another parameter instead of the penetration depth, which does not depend on fluctuations in the development of the cascade process.

The optimum parameter, from our point of view, is the rate of the cascade development.

The energy of the primary particle is spent on the new particle production and ionization losses. At the beginning of the cascade, the rate of increase in the number of particles in a shower is the highest. Then it gradually decreases to zero at the maximum of the shower curve.

In order to determine the primary energy, it is necessary to determine the number of secondary particles at two observation levels, $Z_1$ and $Z_2$, separated by an absorber layer. Based on the obtained measurements, correlation curves are made – the dependence of the cascade size of $\log N_e(Z_1)$ on the rate of the cascade development $dN = \log N_e(Z_1) - \log N_e(Z_2)$.

Figure 2 shows the size-rate (SR) correlation curves for the same showers as in Figure 1. An air layer of 100g/cm$^2$ thick was considered as an absorber. The negative region $dN$ corresponds to low values of the penetration depth.

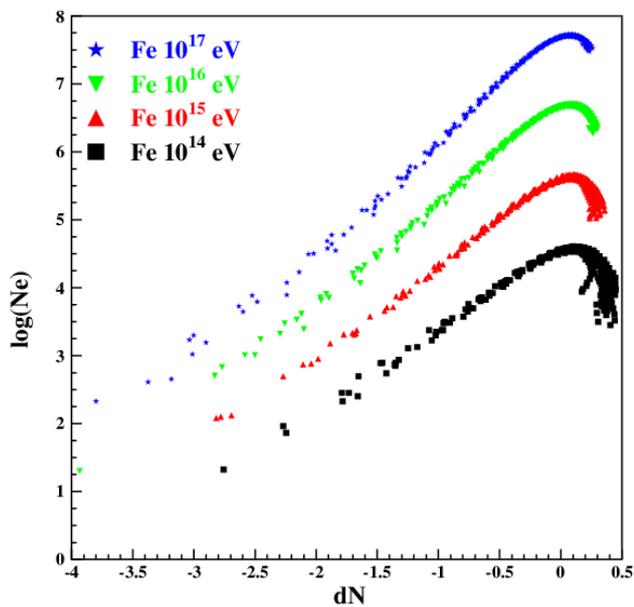

**Figure 2.** The size-rate curves, initiated by iron nuclei of various energies in the atmosphere of Earth. The thickness of the absorber between two observation levels is 100g/cm$^2$.

Figure 2 shows that the SR curves represent a rather ordered pattern. The showers formed by particles of different energies are rather well separated from each other at shallow depths of penetration, in contrast to the cascade curves showed in Figure 1. All SR curves in the negative part of the $dN$ axis are almost parallel to each other and practically do not depend on the depth of the cascade development.

This fact makes it possible to determine the primary energy using an ultrathin calorimeter. The theoretically analyzed absorber with a thickness of 100 g/cm$^2$ of air is about 2.7 radiation units. This corresponds to about 25.5 cm of silicon or 1.5 cm of lead.

The advantage of the presented approach is that the analysis uses a well-measured quantity – the number of particles at the observation level.

In order to study the possibilities of technical implementation of a thin calorimeter based on the presented method, simulation of the development of cascade processes formed by primary particles of various masses and energies was carried out using the GEANT4 software package [19].

### 3. Selection of a calorimeter

At the first stage, we considered a multilayer lead-silicon calorimeter. The layers of *Pb* play the role of dense matter, in which a cascade develops. Silicon is used as a detector.

Lead has a low radiation unit value of $t = 6.4$ g/cm$^2$ and a high density of $\rho = 11.34$ g/cm$^3$. The absorber thickness corresponding to one radiation unit is only 0.56 cm.

The density of silicon is 2.35 g/cm$^3$, and the value of the radiation unit is 22.2 g/cm$^2$. Thus, it is necessary to use silicon with a thickness of 9.4 cm for passage of an absorber in one radiation unit.

In the experiment, the cascade development of in lead is not fixed. It is possible to observe the cascade development only in a detector. Simulation makes this opportunity possible. This is a significant plus.

Taking into account the possibilities of simulation, determinations of the particle characteristics in a detector and an absorber were carried out. The calorimeter size was 50×50 cm. All primary particles fell vertically to the center of the first layer of a calorimeter.

Figure 3 shows the dependence of the number of secondary particles, $N$, in a cascade on the calorimeter thickness $Z$ in a cascade initiated by $10^{12}$ eV iron nuclei in a calorimeter consisting of 30 layers of lead (0.5 cm) alternating with 30 layers (0.5 cm) of silicon.

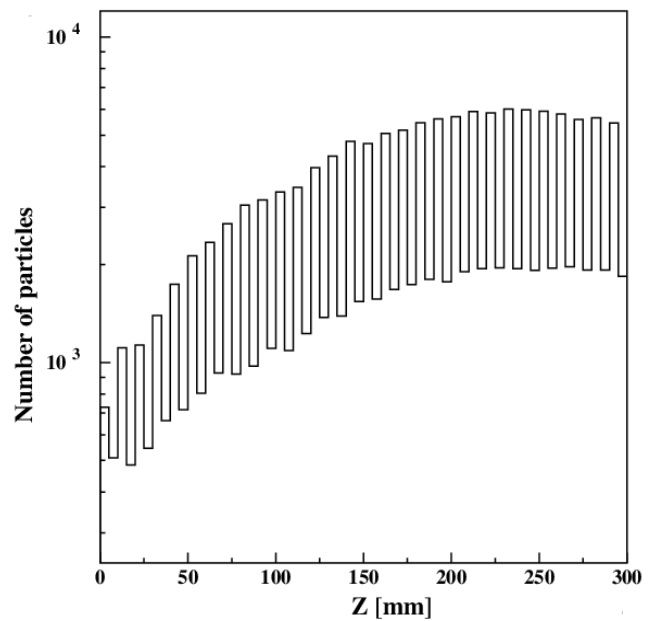

**Figure 3.** The dependence of the number of secondary particles in a cascade on the calorimeter thickness $Z$ during the development of a cascade formed by an iron nucleus with the energy of $10^{12}$ eV in the calorimeter consisting of 30 layers of lead (0.5 cm) alternating with 30 layers (0.5 cm) of silicon.

The presented dependence has a dramatically changing structure with large fluctuations in the number of particles in an absorber and in a detector.

This is due to the significant difference between the value of the critical energy in the material of an absorber and a detector. The main ionization in a calorimeter is generated by relativistic electrons from the developing cascade. The number of electrons in the maximum of the cascade depends on the ratio of the energy of primary particles and the critical energy in a substance.

The critical energy in lead and in silicon varies significantly. It is 7.4 MeV in lead and 37.5 MeV in silicon. Thus, the number of particles in lead and in silicon differs approximately by 5 times.

Thus, when moving from lead to silicon, particles are quickly absorbed and the balance is disturbed.

To avoid transient effects, it is preferable to use a calorimeter in which a detector at level $Z_1$, an absorber between the detectors and a detector at level $Z_2$ would be made of the same material.

The best option, from our point of view, is a calorimeter, which has a unit with a heavy *Pb* layer, leading to the rapid cascade development, and a homogeneous *Si* measuring and absorption unit.

The cascade development in such a calorimeter, consisting of 1 cm of lead and 89 cm of silicon, is shown in Figure 4.

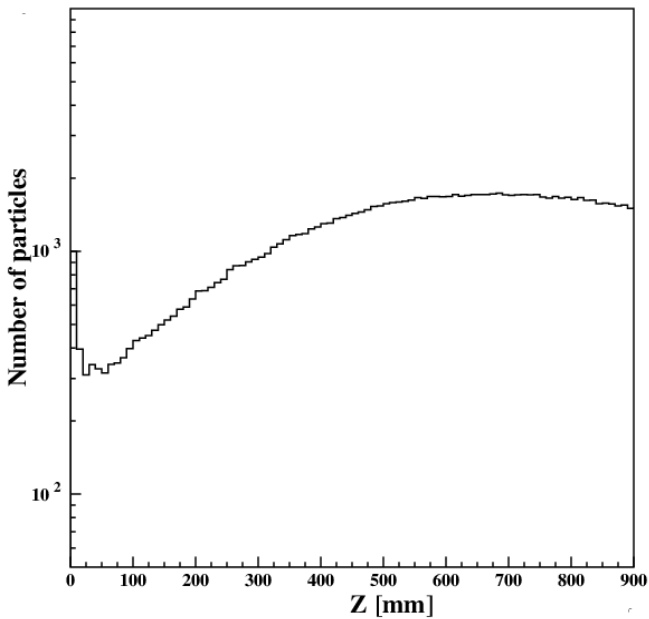

**Figure 4.** The dependence of the number of particles in a cascade on the calorimeter thickness $Z$ during the development of a cascade formed by an iron nucleus with the energy of $10^{12}$ eV in the calorimeter, consisting of 1 cm of lead and 89 cm of silicon.

The presented dependence is a smooth enough curve with a sharp peak in the lead layer region. Consequently, moving away from the *Pb* region, it is possible to carry out measurements at the $Z_1$ and $Z_2$ levels, and reconstruct the energy using the method of SR curves.

### 4. The analysis procedure: the size-rate function

The analysis procedure consisted of several main stages:
1. Simulation of cascades with fixed energies.

The development of 100 cascades formed by the iron nucleus and 100 proton cascades with fixed energies of $10^{12}$ eV and $10^{13}$ eV in the calorimeter, consisting of 1 cm of lead and 89 cm of silicon, were simulated. The dependences of the number of particles in a cascade on the calorimeter thickness are presented in Figure 5.

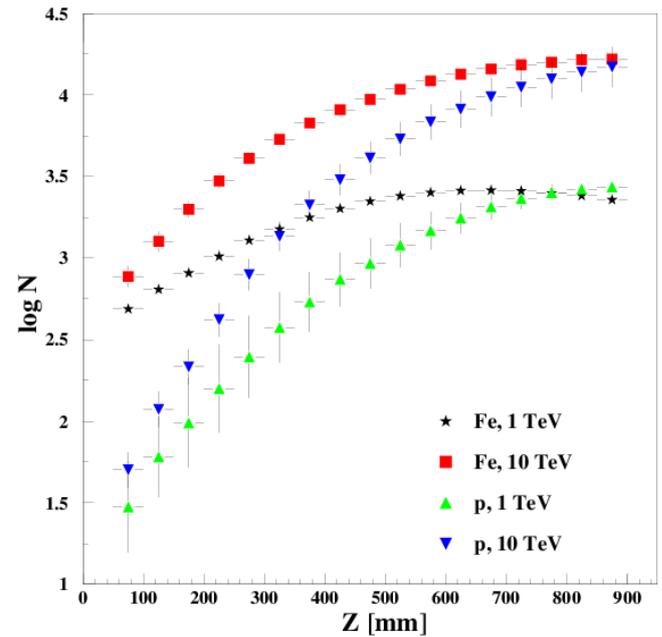

**Figure 5.** The dependence of the number of particles in a cascade on the calorimeter thickness $Z$ during the development of a cascade formed by protons and Fe nuclei with energies of $10^{12}$ and $10^{13}$ eV in the calorimeter, consisting of 1 cm of lead and 89 cm of silicon.

Figure 5 shows that the development of the proton and Fe cascades is significantly different. At low values of the thickness $Z$, for example, at $Z = 100$ mm, Fe showers with an energy of $10^{12}$ eV have log $N$ higher than proton showers with an energy of $10^{13}$ eV.

The most significant factors determining the observed difference are the penetration depth before the first interaction and the number of secondary particles formed when interacting with a substance. Since the average free path is inversely proportional to the cross section before the interaction, the average depth before the first interaction for iron nuclei is substantially less than for protons. And the average multiplicity is significantly lower in the interactions of the proton with a substance.

2. Plotting $logN$-$dN$ – distributions.

Figure 6 shows the average size-rate dependences for the same proton and Fe cascades as in Figure 5.

The absorber thickness between the $Z_1$ and $Z_2$ levels is 10 cm of silicon, that is, about 1 radiation unit.

Figure 6 shows that the size-rate dependences are an ordered structure depending on the primary energy and are practically independent of the type of the primary nucleus.

This fact can also be attributed to the advantages of the presented approach.

3. Fitting third-order polynomial functions for each fixed energy as a function of *logN* from *dN* in the form of:

$$logN (dN) = a_0 + a_1 dN + a_2 dN^2 + a_3 dN^3 \quad (1)$$

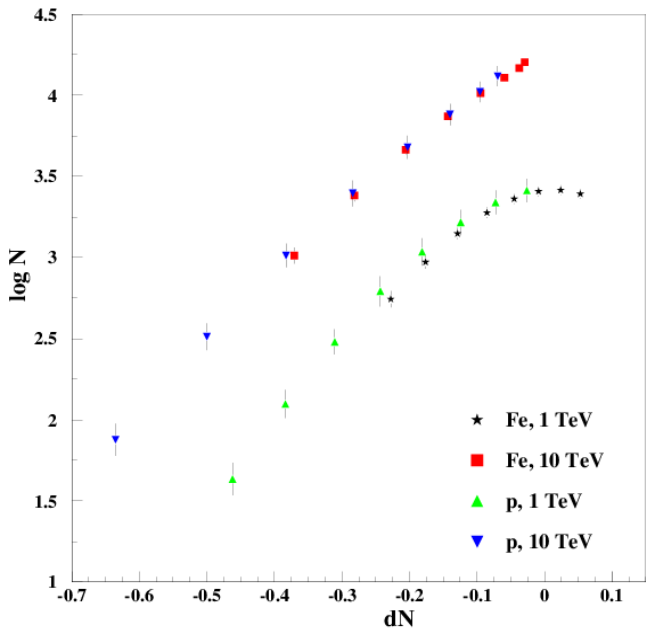

**Figure 6.** The average size-rate dependences for proton and Fe cascades at energies of $10^{12}$ and $10^{13}$ eV in the calorimeter, consisting of 1 cm of lead and 89 cm of silicon. The absorber thickness between two observation levels is 10 cm.

4. Fitting coefficients of $a_0$, $a_1$, $a_2$, $a_3$ depending on energy.

5. Creating the size-rate function: log *N* from *dN* and *E* in the form of:

$$\log N(dN,E) = a_0(E) + a_1(E)dN + a_2(E)dN + a_3(E)dN^3 \quad (2)$$

Using the size-rate function (2) an analysis of test cascades was performed.

## 5. Analysis of test cascades

We considered a calorimeter, consisting of 1 cm of lead and 14 cm of silicon. The first observation level is placed on $Z_1 = 5$ cm. The second observation level was on $Z_2 = 15$ cm.

For the analysis 100 test cascades formed by primary protons and 100 cascades formed by iron nuclei with random energies in the range from $10^{12}$ eV to $10^{13}$ eV were simulated.

Reconstruction of the primary energy was based on dependence (2). In order to determine the energy of the *i*-th test cascade using (2), it is necessary to substitute into the function (2) the "measured" value of the rate $dN_m = \log N_m(Z_1) - \log N_m(Z_2)$ and to vary *E* in order to minimize the difference between "measured" value of the size $\log N_m(Z_1)$ and the size-rate function (2).

Figure 7 shows the obtained accuracies for the energy reconstruction by this procedure.

Errors of energy reconstruction, which is calculated as $\Delta \lg(E_{rec}) = \log(E_{rec}) - \log(E)$, are practically independent of the primary energy.

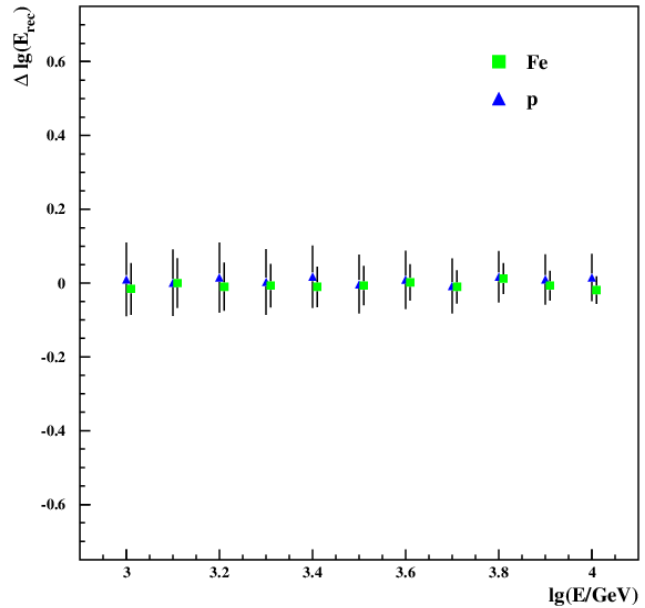

**Figure 7.** Accuracy in energy reconstruction for proton and Fe cascades at different energies in the calorimeter, consisting of 1 cm of lead and 14 cm of silicon.

## 6. Conclusion

An approach for measuring energy of cosmic-ray particles with energies $E \geq 10^{12}$ eV using an ultrathin calorimeter is presented. The method is based on the analysis of the correlation dependence of the cascade size on the rate of development of the cascade process. In order to determine the primary energy, measurements are made based on the number of secondary particles in the cascade, $N_e$, at two observation levels $Z_1$ and $Z_2$, separated by an absorber layer. The correlation curves are almost parallel to each other and practically do not depend on the depth of the cascade development. It makes it possible to determine the primary energy using an ultrathin calorimeter. The best option for applying the method is a calorimeter, which has a unit with a heavy target, leading to the rapid development of the cascade, and a homogeneous measuring and absorption block.

The presented results are only the first step towards applying the SR curves for experimental research.

Subsequently, for a possible technical implementation of the project with an ultrathin calorimeter, it will be necessary to solve a number of issues related to assessment of the effect of various hadron interaction models, a detailed analysis of the effectiveness of the presented approach depending on the energy and mass of the primary nucleus, coordinates and angle of incidence of the primary particle, the choice of the optimum size of a calorimeter and the optimum size of an absorber, a calculation of the response of the actual installation, etc.

However, at present, a fundamental result has been obtained: based on computer simulation, correlation parameters have been selected, making it possible to determine the characteristics of the primary nucleus on the ascending branch of the cascade curve at low values of the penetration depth.

**Acknowledgements**

This paper was supported by a grant from the Ministry of Education and Science of the Republic of Kazakhstan No. AP08855403